\documentclass[12pt,preprint]{aastex}

\begin{document}

\title{The Dependence of Prestellar Core Mass Distributions on the
    Structure of the Parental Cloud}

\author{Antonio Parravano\altaffilmark{1,2},
    N\'estor S\'anchez\altaffilmark{3}, 
    Emilio J. Alfaro\altaffilmark{4}}

\altaffiltext{1}{Universidad de Los Andes, Centro De F\'{\i}sica
         Fundamental, M\'erida, Venezuela}
\altaffiltext{2}{Instituto de Estudios Sociales Avanzados (IESA-CSIC),
         C\'ordoba, Spain} 
\altaffiltext{3}{S.D. Astronom\'ia y Geodesia, Fac. CC. Matem\'aticas,
         Universidad Complutense de Madrid, Spain} 
\altaffiltext{4}{Instituto de Astrof\'isica de Andaluc\'ia (CSIC),
         Granada, Spain}

\newcommand{\beq}{\begin{equation}}
\newcommand{\eeq}{\end{equation}}
\newcommand{\beqa}{\begin{eqnarray}}
\newcommand{\eeqa}{\end{eqnarray}}
\newcommand{\avg}[1]{{\langle #1 \rangle}}
\newcommand{\dis}{\displaystyle}
\newcommand{\e}	{$^{-1}$}
\newcommand{\ee}{$^{-2}$}
\newcommand{\eee}{$^{-3}$}

\def\bd{{\rm bd}}
\def\fnbd{F_{n,\,\rm bd}}
\def\calm{{\cal M}}
\def\mtil{\tilde m}
\def\mpk{m_{\rm peak}}
\def\Rtil{\tilde R}
\def\emch{m_{\rm ch}}
\def\emj{m_{\rm J}^{\rm 0}}
\def\emsun{M_\odot}
\def\msun{$M_\odot \,$}
\def\mmax{$m_{max}$}
\def\emmax{m_{max}}

\slugcomment{\small\it The Astrophysical Journal: accepted}

\begin{abstract}
The mass distribution of prestellar cores is obtained for clouds with
arbitrary internal mass distributions using a selection criterion
based on the thermal and turbulent Jeans mass and applied
hierarchically from small to large scales. We have checked this
methodology comparing our results for a lognormal density PDF with
the theoretical CMF derived by Hennebelle \& Chabrier, namely a
power-law at large scales and a log-normal cutoff at low scales, but
our method can be applied to any mass distributions representing a
star-forming cloud. This methodology enables us to connect the
parental cloud structure with the mass distribution of the cores and
their spatial distribution, providing an efficient tool for
investigating the physical properties of the molecular clouds that
give rise to the prestellar core distributions observed. Simulated
fBm clouds with the Hurst exponent close to the value $H=1/3$ give the
best agreement with the theoretical CMF derived by Hennebelle \&
Chabrier and Chabrier's system IMF. Likewise, the spatial
distribution of the cores derived from our methodology show a surface
density of companions compatible with those observed in Trapezium and
Ophiucus star-forming regions. This method also allows us to analyze
the properties of the mass distribution of cores for different
realizations. We found that the variations in the number of cores
formed in different realizations of fBm clouds (with the same Hurst
exponent) are much larger than the expected root ${\cal N}$
statistical fluctuations, increasing with $H$.
\end{abstract}

\keywords{Stars: formation ---
          Stars: luminosity function, mass function ---
          ISM: evolution  --- 
          ISM: structure}

\section{Introduction}

Is the stellar Initial Mass Function (IMF) universal? This question
has been in the literature for a long time, and is now extended to the
core mass function (CMF) since a close relation between the IMF and
the CMF has been recognized (Motte et al. 1998; Testi \& Sargent 1998;
Alves et al. 2007; Chabrier \& Hennebelle 2010; Michel et al. 2011).
Compared to the CMF, the mass function of stellar systems seems to be
shifted to lower masses by a factor that does not depend on the core
mass. The currently favored conversion efficiency value of the
progenitor core mass to the stellar system is $\sim 1/3$. However,
the origin of this conversion efficiency is still controversial
\citep{Ada96,Mat00,Eno08,Dib11}. Significant variations
in the mass function of young clusters
are observed in the disk of the Galaxy \citep[e.g.,][]{Sca98},
but most of these variations are consistent with random sampling
from a universal IMF \citep{Elm97,Elm99,Kro02,Bas10,Par11}.
The non-lineal processes involved in the star
formation process determine on the one hand the universal
form of the IMF and on the other hand the range of expected variations
of the mass function around this universal form. These fluctuations
arise naturally in IMF models based on deterministic chaos \citep{San99}
and are also observed in three-dimensional hydrodynamic simulations.
Recently \citet{Gir11} performed a parameter study 
of the fragmentation properties of collapsing
isothermal gas cores with different initial conditions and showed that
the density profile strongly determines the number of formed stars,
the onset of star formation, the stellar mass distribution, and the
spatial stellar distribution. Furthermore, the random setup of the
turbulent velocity field in SPH simulation has a major impact in the
different morphology of the filamentary structure, and consequently on
the number of sink particles \citep[as shown by][]{Gir11,Gir12a,Gir12b}.

The ever increasing resolution of magnetohydrodynamic numerical
simulations will provide the answer to many of these questions
\citep{Elm11}. Nevertheless, theoretical IMF models, such as those
proposed by \citet{Pad97,Pad02} or \citet{Hen08,Hen09},
provide analytical solutions that help elucidate the
contribution of the various physical processes involved, but, to
obtain these analytic solutions it is necessary to adopt a series of
assumptions that limit their application to specific cases. The
predictions of these theories have been compared to the numerical data
from simulations. In particular, \citet{Pad04} and \citet{Pad07}
compared the analytical solutions of their theory to
numerical simulations. \citet{Sch10} have also compared the
results from their simulations to the predictions of these theories
and have shown how the clump mass distribution depends on the
turbulence driving mechanism. In between these two approaches are the
phenomenological models, such as the one presented here, that allow
one to address some of the questions stated above, in particular, that
of sensitivity to the initial conditions. The method consists of a
selection criterion based on the thermal and turbulent Jeans mass
which is applied hierarchically from small to large scales. 

\subsection{Aim of the Paper}

The methodology proposed in this work enables a direct connection
between the structure of molecular clouds and the distributions of
generated cores in both mass and space. Thus our first aim is to check
that the results obtained using our methodology are consistent with
those obtained using other methods proposed in the literature and that
have produced reliable results. In particular, we will compare our
results for a lognormal density PDF with the theoretical CMF derived
by \citet{Hen08}, but using two different spatial
distributions of the cloud mass: a) a ramdon cloud, and b) what we
have called a ``corner" distribution where the voxel mass decreases
with the distance to a preselected corner. This exercise allows us to
evaluate the virtue of the method and how the geometry of the cloud
defines the dependence of the standard deviation of the lognormal
density PDF with the smoothing scale R. We have chosen these two very
different spatial structures so as to make it clear how the analytic
formulation of HC08 and the phenomenology presented here are connected
through the scale dependence of the density PDF.

Second, we will explore the formation of cores for different parent
clouds, but considering that the geometry that best describes the
spatial structure of the clouds is fractal. Observations of close
star-forming clouds indicate that the mass distribution in them can be
described as having a fractal structure \citep{Fal92,San05}.
The analysis will be carried out for fractional
Brownian motion (fBm) clouds with a wide range of fractal
dimensions. Specifically, we will focus on the comparative analysis of
the following properties: a) Empirical dependence of density PDF on
the smoothing scale; b) Mass distribution of the cores; c) Spatial
distribution of the generated cores as measured by the surface density
of companions, and, d) Cloud core properties averaged over several
realizations for each $H$.

The paper is organized into five sections, this introduction being the
first. $\S$2 describes the method and defines the main physical
variables of the problem and their range of values in our
simulations. In $\S$3 we check the virtue of our methodology in
reproducing the CMF derived analytically by Hennebelle \& Chabrier
(2008), and in $\S$4, we show the application of this methodology to
fractal clouds generated as fBm clouds with different Hurst exponents
and compare the results with previous approaches to the same physical
systems. Finally, $\S$5 is devoted to summarizing the main
conclusions.

\section{A Discrete Method for a Hierarchical Collapsing Sequence (HCS Method)}

Following the nomenclature in \citet{Mck03} we define a
star-forming clump as a massive region of molecular gas out of which a
star cluster is forming; a core is a region of molecular gas that will
form a single star (or a multiple-star system such as a binary). The
resolution at which a distribution of matter is described can be
limited by the procedure used to generate or measure the distribution,
the capacity of storage of information, or simply can be chosen to
meet a given level of description. In our case, the distribution of
matter is given in a three-dimensional lattice cube of length $L$ with
$N_{vox}^3$ identical cubic voxels. The volume associated with each
voxel is $l_{vox}^3$, where $l_{vox}=L/N_{vox}$. The mass of gas
contained in a voxel centered at coordinates ${\vec r}= l_{vox}
\times [i$\^x$,j$\^y$,k$\^z$]$ is denoted as $m_{i,j,k}$, where $i,j$
and $k$ run from 1 to $N_{vox}$. 

If we assume that the densest voxel contains a mass $\emmax$, and that
the physical conditions in that voxel are such that it is
gravitationally unstable \footnote{\citet{Lar81} noted that ``if
there is a minimum size of bound condensations produced by
supersonic compression processes, this may lead to a lower limit of
the stellar masses".}, (i.e. the Jeans length equals $l_{vox}$),
then the thermal Jeans mass $M_{J,th}$ ($\propto\rho^{-1/2}$) for a
larger cube of $d^3$ voxels is \beq
M_{J,th}=\frac{\emmax^{3/2}}{\sqrt{m_{cube}/d^3}},
\label{eq:mjeanscube}
\eeq where $m_{cube}=\sum\limits_{i,j,k \, \in \, \upsilon}m_{i,j,k}$
is the mass contained in the volume $\upsilon$ whose shape is a cube
of side $d \times l_{vox}$. Note that we are implicitly assuming that
the temperature and the molecular weight are the same in the $d^3$
voxels in the cube.

The turbulent Jeans mass can be expressed in terms of the thermal
Jeans mass as \beq M_{J,turb}=M_{J,th} \frac{V_0^3}{3^{3/2} C_s^3}
\left(\frac{R}{1 {\rm pc}}\right)^{3\eta}= M_{J,th} \,
(\frac{d}{d_{eq}})^{3\eta}, \eeq where $d_{eq}$ is the length (in
$l_{vox}$ units) at which the thermal support and the turbulent
support are equal, $V_0 \simeq 1$ km s$^{-1}$ is the turbulent rms
velocity at 1 pc scale and $C_s=\sqrt{\frac{\gamma_g k T}{\mu n_H}}
\simeq 0.22 \left(T/10 {\rm K}\right)^{1/2}
\left(\mu/2.33\right)^{-1/2}$ km/s is the sound speed, where
$\gamma_g$ is the adiabatic index and $\mu$ is the molecular weight.
The exponent $\eta$ is the exponent of the Larson's (1981) velocity
dispersion versus size relation and is related to the 3D power
spectrum index of the velocity field $\eta=(n-3)/2$ where $n=11/3$ for
the Kolmogorov case and 4 for the Burgers case. \citet{Kri07}
estimate $n\sim 3.8-3.9$ ($\eta\sim 0.4-0.45$) from high resolution
hydrodynamic simulations of isothermal supersonic turbulence, in
agreement with \citet{Sch09} who estimate $\eta\sim 0.45$ from
simulations of supersonic isothermal turbulence driven by mostly
compressive large-scale forcing. \citet{Fed10} showed that
$\eta$ depends on the nature of the turbulence forcing mechanism; 0.43
for solenoidal forcing and 0.47 for compressive forcing.

\citet{Mye92} first included both thermal and nonthermal
motions in a model of star formation in dense cores. \citet{Mck03}
focused on the nonthermal part in their turbulent core model
for massive star formation, but then showed how it is possible to
smoothly join on to the thermal Jeans mass. Following
\citet[][hereafter HC08]{Hen08}, who explicitly included both thermal
and nonthermal motions, the Jeans mass can be expressed as \beq
M_{J}=M_{J,th} \left\{1 +
\left(\frac{d}{d_{eq}}\right)^{2\eta}\right\}^{3/2}.
\label{eq:mjeans_therm*turb}
\eeq

To apply the Jeans criteria in eq. (\ref{eq:mjeans_therm*turb}) to any
cube in the array it is only necessary to know $m_{i,j,k}$ for all
$i,j,k$ in the array and the parameters $\emmax$ and $d_{eq}$. Note
that the physical size $l_{vox}$ of the voxels is not needed to
determine whether a given cube is Jeans unstable. However, as shown
below, the parameter $d_{eq}$ depends on the physical conditions that
determine $l_{vox}$ and on the increase of the velocity dispersion
with distance. 

We propose here a procedure to obtain the prestellar core mass
distribution which is based on a hierarchical collapsing sequence
(hereafter the HCS method) in which the densest regions collapse first
and form the smaller objects. At small scales, thermal support
dominates and determines the core mass distribution at low masses,
whereas, at the largest scales turbulence dominates the support and
determines the mass distribution at high masses, as in the analytical
theory of the IMF proposed by HC08. The discrete mass distribution
$m_{i,j,k}$ is checked at all scales starting with the smallest, that
is at the scale of one voxel, i.e. $d=1$. By construction only the
densest voxel (the one with mass $\emmax$) is marginally unstable
under the thermal Jeans criterion, however the small turbulent support
is enough to suppress the collapse at one voxel scale. Then, cubes of
side $d=2$ are checked to find those that fulfill the condition
$m_{cube} \ge M_{J}$. In the cubes fulfilling this condition, a core
of mass $m_{core}=M_{J}$ is assumed to form giving rise to a stellar
system of mass $m_{*}=\epsilon \, m_{core}$. The remaining gas
$m_{cube}-\epsilon \, M_{J}$ is assumed to become inactive. After
this, cubes of side $d=3,4,...,N_{vox}/2$ are considered
consecutively. Note that the properties of the velocity field are not
considered explicitly, but are taken into account implicitly by means
of the parameters $d_{eq}$ and $\eta$.

To fix the voxel length $l_{vox}$ we use the assumption that the mass
$\emmax$ contained in the densest voxel is gravitationally marginally
stable. The radius of a Bonnor-Ebert sphere is $R_{BE} \simeq
0.486\,R_G$ and the thermal Jeans length is $l_{J,th}=\sqrt{\pi}
\,R_G$, where $R_G$, the gravitational length (McKee \& Ostriker
2007), is \beq R_G=\sigma_{th}/(G\,\rho)^{1/2} \simeq 0.21 \left(
\frac{T}{10 {\rm K}}\right)^{1/2} \left(
\frac{\mu}{2.33}\right)^{-1/2} \left( \frac{n_H}{10^4
 cm^{-3}}\right)^{-1/2},
\label{eq:RG}
\eeq and where $n_H$ is the hydrogen nucleus number density. 

Since the density in the densest voxel is $n_H \simeq 41 \left(
\frac{\emmax}{1 \, \emsun}\right)/\left( \frac{l_{vox}}{1 {\rm
 pc}}\right)^3$ cm$^{-3}$, the voxel length is \beq l_{vox}= l_1
\left( \frac{\emmax}{1 {\emsun}}\right) \left( \frac{10 {\rm K}}{T}
\frac{\mu}{2.33}\right)\,\, {\rm pc},
\label{eq:lvox}
\eeq where $l_1=0.15$ if $l_{vox}^3=\frac{4}{3} \pi R_{BE}^3$. Note
that this value is close to the value $l_1=0.18$ obtained if
$l_{vox}^3=\frac{4}{3} \pi \left( \frac{l_{J,th}}{4}\right)^3$. We
adopt $l_1=0.15$ here.\footnote{ If we consider that the smallest
 single brown dwarf that can be formed has a mass of the order of
 0.02 \msun and the progenitor core is about three times this mass,
 then $\emmax \thicksim 0.06 \emsun$. For this value of $\emmax$ and
 the fiducial values $T=10$ K and $\mu=2.33$, eq. (\ref{eq:lvox})
 gives $l_{vox}=0.009$ pc, a value that roughly agrees with the value
 predicted by Larson's law for this mass \citep{Lar81,Kau10}.}

Finally, the parameter $d_{eq}$ depends on the turbulent rms velocity
which is assumed to increase with the size $R$ of the region following
the Larson relation $\langle V_{\rm rms}^2 \rangle = V_0^2 \times
\left(\frac{R}{1{\rm pc}} \right)^{2\eta} $, where $V_0 \simeq 1$ km
s$^{-1}$. The value $R_{eq}$ at which thermal and turbulent support
are equal can be expressed in terms of the sound speed $C_s$ as
$R_{eq}/{1{\rm pc}}= \left(\sqrt{3} C_s/V_0 \right)^{\frac{1}{\eta}}$.
Therefore, in terms of the Mach number at 1 pc scale ${\calm_{\rm 1
 pc}}=V_0/C_s$, the number of voxels $d_{eq}$ for which thermal and
turbulent support are equal is \beq d_{eq}\equiv
R_{eq}/l_{vox}=\left(\sqrt{3}/{\calm_{\rm 1
 pc}}\right)^{\frac{1}{\eta}}/(l_{vox}/{1{\rm pc}}).
\label{eq:d_eq}
\eeq

We apply the HCS procedure first to mass distributions with a
log-normal density probability distribution function (PDF) in order to
compare our numerical results to the analytical CMFs derived by
HC08. Later the procedure is applied to fractal clouds with density
PDFs that are not necessarily log-normal.

\section{Comparison to the HC08 Analytical CMFs Theory}
The analytical theory for the IMF developed in HC08 is based on an
extension of the \citet{Pre74} statistical formalism
applied in cosmology. When applied to the mass function of molecular
cloud cores, the original Press-Schechter formalism has the problem
that structures inside structures are not counted. This
“cloud-in-cloud” problem was overcome by assuming a conditional
probability of finding a collapsed region of mass scale $M$ inside a
collapsed region of mass scale $M'$
\citep[][and references therein]{Inu01}.
Additionally, in the Press-Schechter theory the structures
are identified with over-densities in a random field of density
fluctuations; i.e. a normal distribution in density. Instead, HC08
assume a log-normal distribution in density, as suggested by numerical
simulations of non-self-gravitating supersonic isothermal turbulence
\citep{Vaz94,Pad97,Pas98,Ost01,Kri07} and observations
\citep{Kai09,Kai11}. Finally, HC08 assume that at any
smoothing scale $R$ the mass distribution in the cloud is such that
the density PDF is always log-normal but with a standard deviation
$\sigma(R)$ that decreases with $R$ as \beq \sigma^2(R)=\sigma_0^2
\left(1-\left(\frac{R}{L_i}\right)^{2\eta}\right),
\label{sig2_R}
\eeq where $L_i$ is the injection scale and $\sigma_0$ is the width of
the density distribution at maximum resolution $R \ll L_i$. The
density PDF at resolution $R$ is then \beq {\it
 P}_R(\delta,R)=\frac{1}{\sqrt{2\pi \sigma^2(R)}}
exp[-\frac{[\delta-\frac{\sigma^2(R)}{2}]^2}{2 \sigma^2(R)}],
\label{pdf_sig2_R}
\eeq where $\delta=ln(\rho/\bar \rho)$ and ${\bar \rho}$ is the cloud
mean density. For this scale-dependent density PDF their theory
identifies gravitationally-bound prestellar cores with regions that
have a density threshold given by the requirement that a fluctuation
contains at least one local (thermal or turbulent) Jeans mass. As
before, the turbulent rms velocity is assumed to correlate with size
$R$ following the Larson power-law $\langle V_{\rm rms}^2 \rangle =
V_0^2 \times \left(\frac{R}{1{\rm pc}} \right)^{2\eta}$. The places
where the average density at scale $R$ is larger than the density
threshold contain more than one Jeans mass and are expected to form
prestellar cores of mass smaller than or equal to the mass contained
in that region. This is because at smaller scales it may happen that
the region is not uniform but composed of smaller, denser regions
embedded into a more diffuse medium. If these denser regions contain
one Jeans mass, the end product of the collapse is likely to be a
cluster of objects whose mass is close to the mass of the
smaller/denser regions and not to the mass in the volume at scale $R$.
Taking into account the probability of finding these unstable
sub-structures, HC08 express their core mass function as:

\beq \psi_{\rm HC}(\mtil)\equiv \frac{1}{{\cal N}_{tot}}\frac{d{\cal
 N}}{d\ln \mtil} \propto
\frac{1}{(\mtil\Rtil^3)^{1/2}}\left[\frac{1+(1-\eta)\calm_*^2\Rtil^{2\eta}}
 {1+(2\eta
 +1)\calm_*^2\Rtil^{2\eta}}\right]\exp\left\{-\frac{[\ln(\mtil/\Rtil^3)]^2}
 {2\sigma_0^2}-\frac{\sigma_0^2}{8}\right\},
\label{psiHC08}
\eeq where \beq \mtil\equiv \frac{m}{\emj}=
\Rtil(1+\calm_*^2\Rtil^{2\eta}),
\label{norm-psiHC08}
\eeq $\emj$ is the Jeans mass at the average cloud density, $\Rtil$ is
the radius of the clump in units of the Jeans length at the average
cloud density, $\calm_*$ is the characteristic Mach number at the
Jeans scale, and $\eta$ ($\sim 0.4-0.45$) is the exponent of the
linewidth-size relation. The $\psi_{\rm HC}(\mtil)$ mass distribution
in eqs. (\ref{psiHC08} - \ref{norm-psiHC08}) represents the stellar
IMF, whereas $\psi_{\rm HC}(m)$ represents the CMF.

At low masses ($\mtil <1$), the form of $\psi_{\rm HC}$ is log-normal.
At moderately high masses ($m\ga \emj$) the IMF approaches the power
law \beq \psi_{\rm
 HC}\propto\mtil^{-(\eta+2)/(2\eta+1)}\equiv\mtil^{-\Gamma_{\rm
 HC}}~~~~~(\mtil\ga 1), \eeq which gives $\Gamma_{\rm HC}\simeq
1.3$ for the value $\eta\simeq 0.4$ they adopt, in agreement with the
\citet{Sal55} value. At very high masses ($\mtil \gg 1$) the IMF
drops off more steeply with mass, becoming a log-normal type
distribution again. Their results for a non-isothermal equation of
state are considerably more complicated, but they are qualitatively
consistent with the isothermal theory \citep{Hen09}.

Note that only the scale-dependent log-normal PDF of the density and
the properties of turbulence are considered in the HC08 theory. The
detailed spatial distribution of matter is irrelevant, even when,
implicitly, their results refer to the kind of gas distributions
associated with turbulence. Instead, the HCS method proposed here
explicitly takes into account the spatial distribution of the matter,
and as shown below, the resulting core mass distribution is sensitive
to this distribution. 

For the moment we do not consider the radiative feedback
\citep{Hol99,Gor02,Kru07,Kru11,Bat09,Bat12,Pri09}, but it is
expected that its efficiency also greatly depends on the spatial
distribution of the matter.

\subsection{Artificial mass distributions with log-normal density
PDFs: ``corner" and ``Random" clouds.}

To compare with HC08 we first consider two extreme mass distributions
that are useful for showing the importance of the spatial structure of
the cloud on the resulting mass function of collapsing cores. The
first is a ``random cloud" in which the masses $m(i,j,k)$ and the
positions $(i,j,k)$ are uncorrelated. The second is a ``corner cloud"
in which the locations of the mass voxels $m(i,j,k)$ are ordered in
such a way that the density decreases as the sum $i+j+k$ increases. 

\begin{figure}
\epsscale{0.9} \plotone{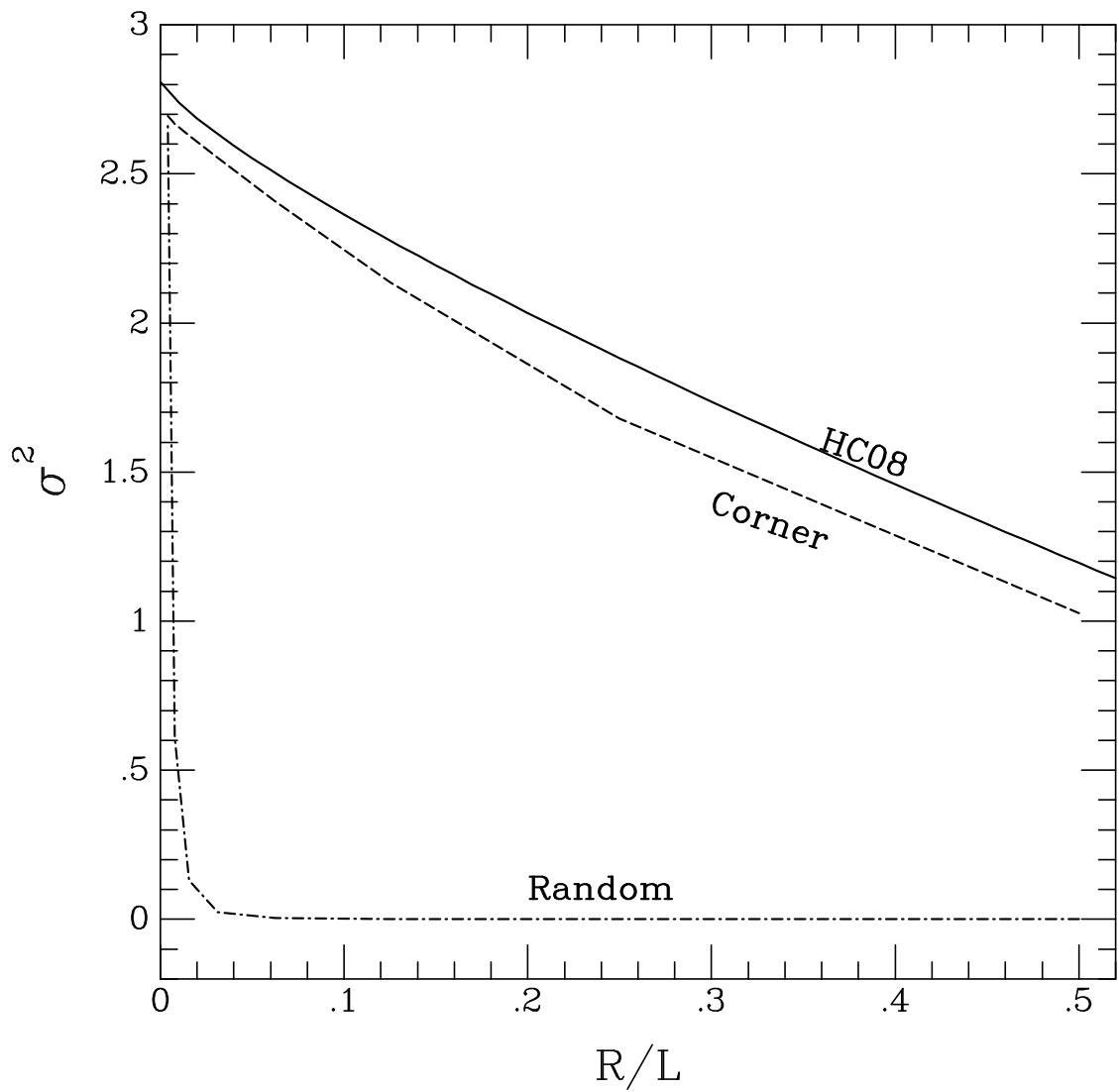}
\caption
{Dependence of the standard deviation on the smoothing scale for two
 different mass distributions and the dependence assumed by HC08 in
 eq. (\ref{sig2_R}) with $L_i=L$ and $\sigma_0$ calculated from
 eqs. (\ref{mach-cloud} - \ref{s2-cloud}). The parameters in the
 three cases are $N_{vox}=2^8$, $\emmax=0.07 \emsun$, $T=10$K,
 $\mu=2.33$ and $\eta=0.4$, corresponding to the third case in Table
 1. }
\label{s_VS_R_c_r}
\end{figure}

Since the lattice is regular and the voxel size $l_{vox}$ is fixed by
$\emmax$, the mass distribution of the voxels (that is the number of
voxels with a given mass) is proportional to the density PDF, and
therefore the standard deviation $\sigma_0$ is the same for the PDF of
the masses $m(i,j,k)$ and for the PDF of the gas density, and has the
same meaning as in eq. (\ref{psiHC08}). The PDFs of the masses
$m(i,j,k)$ in both types of clouds follow the same log-normal function
\beq dN/dln m= \frac{N_{vox}^3}{\sigma_0 \sqrt{2\pi}}
exp[-\frac{[ln(m/\bar{m})-\frac{\sigma_0^2}{2}]^2}{2 {\sigma_0}^2}],
\label{eq:lognormal}
\eeq where $\bar{m}$ is the mean mass per voxel and the peak of the
distribution occurs at $ln(m_0)=ln(\bar{m})-\sigma_0^2/2$. However,
the dependence of the standard deviation $\sigma(R)$ on the smoothing
scale $R$ is, as shown in Fig. \ref{s_VS_R_c_r}, very different. The
smooth gradients in the mass distribution of the ``corner cloud"
produce a $\sigma(R)$ that is close to the HC08 dependence in
eq. (\ref{sig2_R}). In the ``random cloud" case, $\sigma(R)$ rapidly
drops to zero due to the unphysical discontinuous densities that make
the average density in any volume containing a relatively small number
of voxels very close to $\bar{\rho}$.

\subsection{Comparison of HC08 and HCS CMFs for equivalent clouds.}

To compare the core mass function corresponding to a particular mass
distribution $ m_{i,j,k}$ with $\psi_{\rm HC}$ we have to determine
the set of HC08 parameters ($\sigma_0$, $\emj$, $\calm_*$) from the
HCS input parameters ($N_{vox},\emmax,T,\mu,\eta$). 

The voxel size $l_{vox}(\emmax,T,\mu)$ is calculated from
eq. (\ref{eq:lvox}), and the mean cloud density is $\bar{n}_H=\sum
m_{i,j,k}/(m_H \, L^3)$ with $L= N_{vox} \times l_{vox}$. 

Following HC08 \citep[see also][]{Hen09} the Mach number
at the Jeans scale at the cloud mean density $\bar{n_H}$ is \beq
\calm_*^2 \simeq \left(T/10 {\rm K}\right)^{\eta-1}
\left(\bar{n_H}/10^4 {\rm cm}^{-3}\right)^{-\eta} \eeq and the Jeans
mass at the cloud mean density $\bar{n_H}$ is \beq \emj=m_1
\left(T/10 {\rm K}\right)^{3/2}
\left(\mu/2.33\right)^{-3/2}\left(\bar{n_H}/10^4 {\rm
 cm}^{-3}\right)^{-1/2} \,\, \emsun,
\label{m_jeans_mean}
\eeq where the value of $m_1$ depends on the definition of the Jeans
mass. HC08 adopt $m_1\simeq1$, but if $\emj$ is assumed to be the mass
in a sphere of diameter $l_{J,th}=\sqrt{\pi} R_G$ then
$m_1\simeq6.6$. If instead $\emj$ is assumed to be the mass in a
sphere of diameter $l_{J,th}/2$ then $m_1\simeq 0.8$, close to the
value adopted in HC08. We assume $m_1=1.08$, corresponding to the case
in which $\emj$ is assumed to be the mass in a Bonnor-Ebert sphere.

The Mach number at the cloud scale $ \calm =\frac{V_0}{C_s}
(\frac{L}{1{\rm pc}})^{\eta}$ is \beq \calm \simeq (1/0.22) \left(T/10
{\rm K}\right)^{-1/2} \left(\mu/2.33\right)^{1/2} \left(L/1{\rm
 pc}\right)^{\eta},
\label{mach-cloud}
\eeq and the width of the density distribution $\sigma_0$ at the cloud
scale is \beq {\sigma_0}^2=ln(1+b^2 \calm^2),
\label{s2-cloud}
\eeq where $b^2 \approx 0.25$ (see also Hennebelle \& Chabrier 2009).

For a given set of parameters $\sigma_0$ and $\emmax$, the masses in
the $N_{vox}^3$ voxels in the ``random" or the ``corner" clouds can be
assigned following the log-normal distribution in
eq. (\ref{eq:lognormal}), requiring that the mass in the densest voxel
is $\emmax$. Then the resulting core mass function for $m_{i,j,k}$ can
be compared to $\psi_{\rm HC}$. However, what is the appropriate
value for $\emmax$ to make the comparison? We assume that the
appropriate value of $\emmax$ is the value for which the mean cloud
density $\bar{n}_H$ is such that the whole cloud is close to virial
equilibrium; that is, the virial parameter \citep{Lar81,Ber92},
defined as $\alpha_{VIR}=5 \sigma_{vel}^2 R /(G
M_{cl})\simeq 5 (\frac{\Delta v}{2} )^2 (L/2) /(G M_{cl})$ is equal to
one. Therefore, $(M_{cl}/1 \emsun) \simeq 150 (L/1\,{\rm
 pc})^{2\eta+1}$ if $({\Delta v/1\,{\rm km \, s^{-1}})} \simeq
\left(L/1{\rm pc} \right)^{\eta} $. The cloud mass $M_{cl}=\sum
m_{i,j,k}$ depends on the form of the density PDF, on the number of
voxels in the lattice and on $\emmax$. The lattice size $L$ depends on
$\emmax$ through $l_{vox}(\emmax)$ in eq. (\ref{eq:lvox}). 

Table 1 gives the $\emmax$ values that fulfill the above two
conditions when the PDF of the mass in voxels is log-normal. Table 1
also gives various derived quantities and the corresponding HC08
parameters. Note that the virial parameter is calculated omitting the
effect of the pressure produced by the medium surrounding the
cloud. \citet{Kai11} estimate that the pressures
supporting the clumps against dispersal amount in total to about one
third of the pressure driving their dispersal. Therefore, cloud masses
in Table 1 exceed the Jeans masses in eq. \ref{eq:mjeans_therm*turb}
since $M_{J,th}$ is based on the stability of a Bonnor-Ebert sphere. 

\begin{table}
\begin{center}
\begin{tabular}{c}
\hspace{0.0cm}TABLE 1\\
\hspace{0.0cm} Properties of clouds with log-normal density PDFs\\
\end{tabular}\\
\begin{tabular}{rrrrcrcrrrrc}
\hline \hline $N_{vox}$&$T $& $\emmax $& $d_{eq}$& $M_{cl}$& $L$&
${\bar n_H} $& $\sigma_0$ & & $\calm$ &$\calm_*$& $\emj $\\ &$[{\rm
  K}]$& $[\emsun]$&& $[\emsun]$& $[{\rm pc}]$& $[{\rm cm}^{-3}]$&
& & && $[\emsun]$\\ \hline 64& 10& 0.109&  5.67&  156&  1.05&
5616&  1.42 & \vline & 5.09&  1.12&  1.44\\ 128& 10& 0.088&
7.02&  370&  1.69&  3166&  1.53 & \vline & 6.17&  1.26&
1.92\\ 256& 10& 0.070&  8.83&  862&  2.69&  1831&  1.64 &
\vline & 7.43&  1.40&  2.52\\ 256&  8& 0.042&  8.91&  521&
2.02&  2622&  1.64 & \vline & 7.40&  1.40&  1.51\\ 256& 12&
0.106&  8.79&  1301&  3.39&  1375&  1.64 & \vline & 7.44&
1.41&  3.83\\ \hline
\end{tabular}
\end{center}
\end{table}

\begin{figure}
\epsscale{0.9} \plottwo{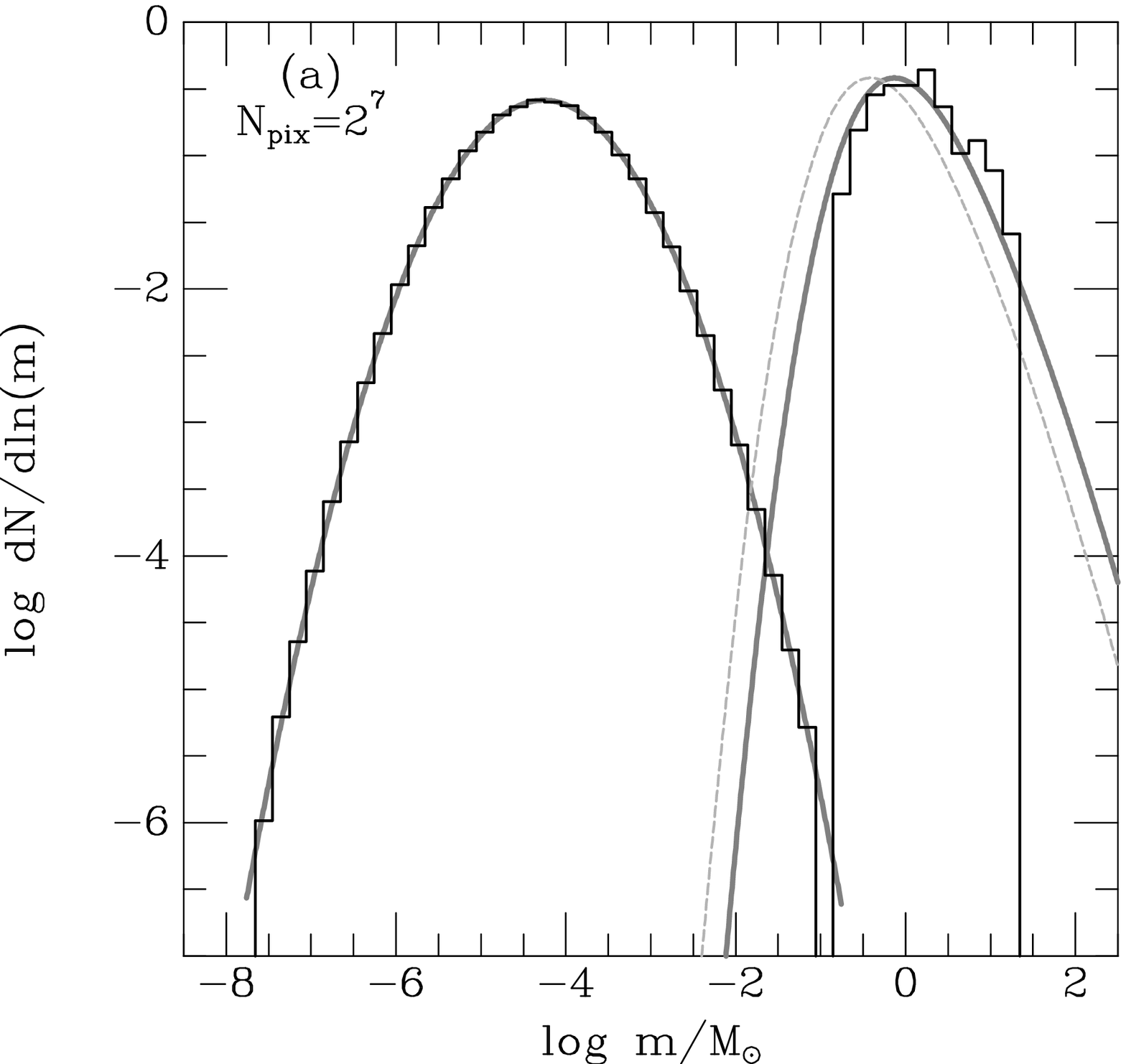}{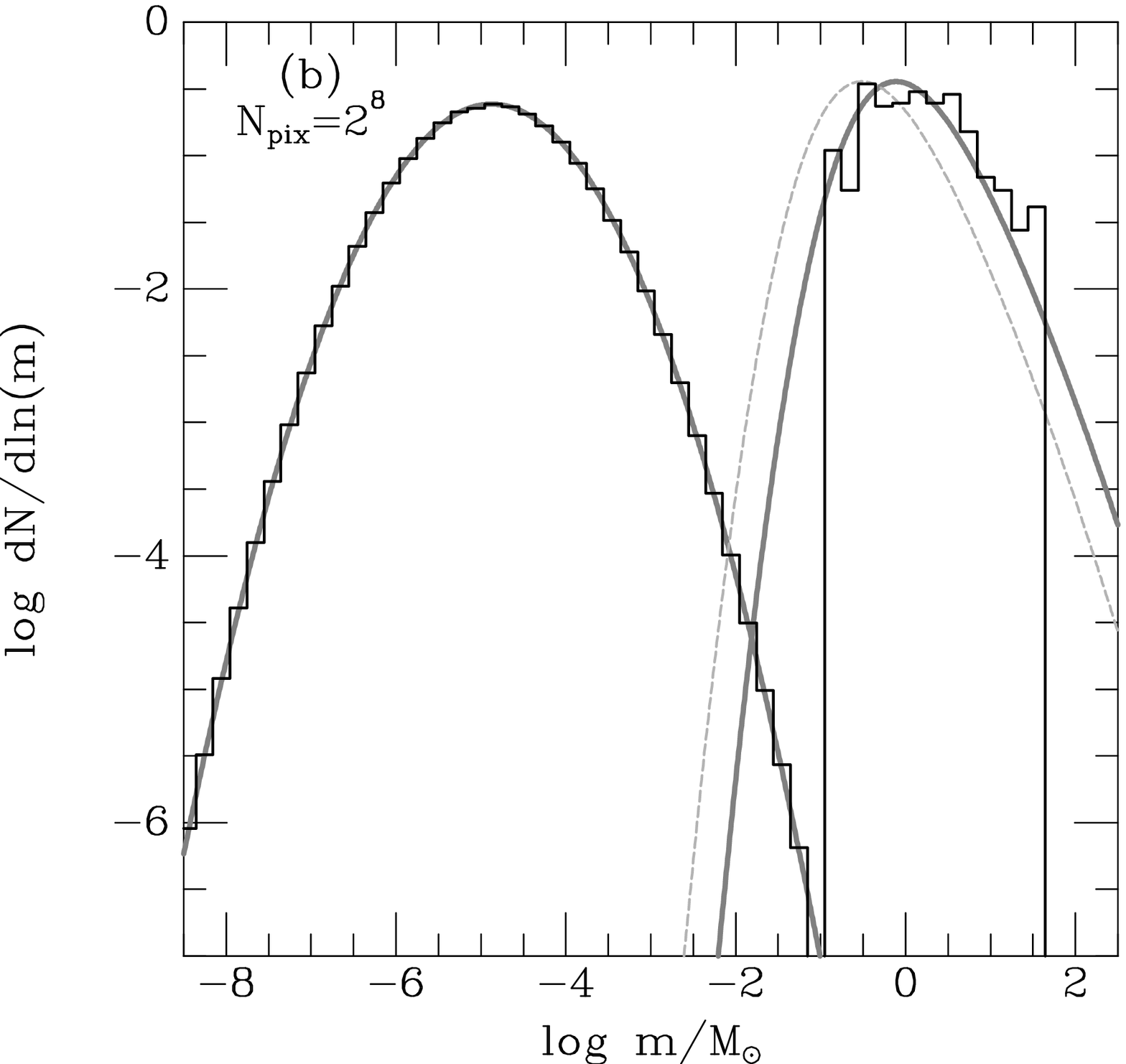}
\caption{Histograms of the PDF of mass in voxels and the resulting
 core mass function for the ``corner cloud" case for two array sizes,
 $N_{vox}=128$ (a) and 256 (b). The parameter values are those in the
 second and third lines of Table 1. The continuous gray curve at the
 left of each panel is a log-normal with the appropriate values
 $\sigma_0$ and $\emmax$. The continuous gray curve at the right of
 each panel is $\psi_{\rm HC}(m)$ and the dashed gray curve is the
 normalized HC08 IMF $\psi_{\rm HC}(m/\emj)$. }
\label{fig:corner-hc08}
\end{figure}

For the ``corner cloud" and two array sizes, $N_{vox}=128$ and 256,
Figure \ref{fig:corner-hc08} shows the histograms of the PDF of voxel
masses $m_{i,j,k}$ and the core masses obtained with the HCS
method. Figure \ref{fig:corner-hc08} also shows the analytical PDF and
core mass function from HC08; the dashed gray curve is $\psi_{\rm HC}$
from eq. (\ref{psiHC08}) as function of $\mtil=m/\emj$, whereas the
continuous gray curve is not normalized to $\emj$. Agreement with the
non-normalized HC08 CMF is good for all cases in Table 1. The rapid
falloff of the core mass function at low $m$ is due to the fact that
by construction we have set the most massive voxel to a mass of about
0.06, and that sets the minimum mass of a core. At high $m$ the rapid
falloff is because the highest mass of a core is limited by the mass
of the cloud that remains after the formation of the smaller cores.
Results very similar to those in Fig. \ref{fig:corner-hc08} are
obtained for spherically symmetric distributions in which the densest
voxel with mass $\emmax$ is located at the center of the array and the
mass of the remaining voxels decrease from the center n accord with
the log-normal PDF. These results indicate that the HCS numerical
method captures the main features of the HC08 analytical
theory. However, the HCS method is not restricted to mass
distributions with density PDFs following
eqs. (\ref{sig2_R}-\ref{pdf_sig2_R}).

HC08 express their stellar IMF in terms of $\emj$ which is about $3 \,
\emsun$ for a mean cloud density ${\bar n_H} \sim 1000 \,{\rm
 cm}^{-3}$. Instead of normalizing masses to $\emj$ we assume that
the non-normalized function $\psi_{\rm HC}(m)$ represents CMF and that
the mass function of stellar systems is shifted to lower masses by a
factor $\epsilon$ of about 1/3. The evolution of the CMF to the IMF
has not been definitively established but magnetically-driven outflows
(Matzner \& McKee 2000) are expected to produce a mass independent
efficiency factor in the range $30-50\%$; we assume a value of
$\epsilon \simeq 1/3$. Note that our procedure for obtaining the
stellar mass function for a particular gas distribution, as well as
the theories of the IMF such as those of \citet{Pad02}
\citep[see also][]{Pad07} and HC08, predict the system IMF, whereas
observations that cannot resolve close binaries determine an effective
IMF since unresolved binaries are counted as single stars with an
effective mass. The determination of the individual star IMF, in
which each star that is a member of a multiple system is counted
separately \citep{Par11}, is beyond the scope of the present
study. The fraction of the cloud mass that eventually becomes
collapsing cores ($\sim 0.6$), as well as the number of these cores,
depends weakly on the gas temperature but the mean core mass is
proportional to $(T/10 {\rm K})^{1+1/\eta}$.

Contrary to the ``corner cloud" or the spherically symmetric
distributions, a ``random cloud" that has the same density PDF does
not produce low mass collapsing cores, showing that the spatial
structure of the cloud is very important not only for the spatial
distribution of the protostellar objects, but also for the efficiency
and mass function of the star forming objects. For the PDF of mass in
voxels in Fig. (\ref{fig:corner-hc08}-b), the lowest mass core that
collapses in the random cloud is $1.3 \emsun\sim 20 \, \emmax$,
whereas for the corner cloud it is $0.12 \emsun \sim 2 \, \emmax $.
If the virial parameter of the cloud is increased to
$\alpha_{VIR}=2.5$ (i.e. $M_{cl}\simeq 1000 \emsun$ and $L\simeq 5$
pc), then the random cloud produces only a couple of high mass cores
($m\sim 100 \emsun$) but the corner cloud still produces a similar
number of cores as in the $\alpha_{VIR}=1$ case.

It is important to point out that these results show that the
dependence of the dispersion of the density PDF on the smoothing scale
strongly affects the resulting mass distribution of collapsing cores.
\citet{Fed10} showed that the density PDF in their simulations
are roughly consistent with log-normal distributions for both
solenoidal and compressive forcings, even when the distributions
clearly exhibit non-Gaussian higher-order moments. However, the
dispersion of the density PDF is highly sensitive to the turbulence
forcing, and therefore they conclude that the theoretical CMF/IMF
derived in HC09 is strongly affected by the assumed turbulence forcing
mechanism. In the following we consider fractal mass distributions
with roughly log-normal distributions, but with adjustable $\sigma(R)$
functions.

\section{Fractal Clouds}

Fractal clouds are known to be a good representation of star forming
regions \citep{San07a,San07b,Elm02,Elm10}. These kinds
of clouds are easy to construct by means of recurrence procedures that
produce hierarchical self-similar mass distributions. Fractal
distributions are observed over a wide range of scales, from dense
cores to giant molecular clouds \citep{Ber07}. In
particular, numerical simulations of supersonic isothermal turbulence
\citep{Kri07,Fed09} showed that the density field has an approximately
fractal structure.

We focus on fractional Brownian motion clouds
\citep[hereafter fBm clouds;][]{Stu98,Elm02,Miv03,San10}
that have been used to represent the internal
structure of molecular clouds. The fBm clouds are generated following
the procedure described in \citet{Miv03}. That is,
a field with Gaussian distribution intensity $\tilde{I}_{i,j,k}$ is
obtained by first filling a lattice in wavenumber space $({\rm
 k}_x,{\rm k}_y,{\rm k}_z)$ with a random phase and Fourier
amplitudes proportional to $\vert \overrightarrow{{\rm
 k}}\vert^{-(H+E/2)}$, where $\vert\overrightarrow{{\rm
 k}}\vert=({{\rm k}_x}^2+{{\rm k}_y}^2+{{\rm k}_z}^2)^{1/2}$, $H$
is the drift (or Hurst) exponent, and $E$ is the dimension of the
lattice. Subsequently, an inverse fast Fourier transform is applied
to generate an intensity distribution $I_{i,j,k}$ in real space with a
Gaussian intensity distribution. Since intensities must be real
values, the Fourier amplitudes and phases in the wave number space
have to match the appropriate symmetry conditions \citep{Stu98}.
The exponent $H$ corresponds to a power spectrum of the
intensity distribution $\gamma = 2H + E$. High resolution numerical
experiments of supersonic isothermal turbulence driven by solenoidal
forcing, obtained by \citet{Fed09}, have been characterized
by a Hurst exponent $H=0.39$, corresponding to a box counting
dimension $D_b\simeq 2.61$. Even when the three dimensional density
and velocity fields are correlated \citep{Miv03},
their relation strongly depends on the physical processes
involved in the gas dynamics, for example on the energy injection
mechanism as shown in \citet{Fed09}. HC09 theory takes the
exponent of the power spectrum of the density and velocity fields into
account, but in their CMF analytical solutions the two exponents are
assumed to be equal. This assumption is used in the previous section
to compare our method with HC09 (i.e. through the exponent $\eta$ in
eq. (\ref{sig2_R})). Note that if the two exponents are assumed to be
equal, $H=1/3$ for a dissipationless cascade of energy through an
incompressible fluid (i.e. $\gamma=11/3$ for $E=3$). In this section,
for the application of our method to fBm clouds, we use the quantities
$\eta$ and $H$ to parameterize separately the velocity and the density
fields; the first parameterize the turbulent support, and the second
determine average properties of the mass distribution in fBm clouds,
including $\sigma(R)$. FBm clouds have a two-point correlation
function of the intensity $G_I(\lambda)\varpropto\lambda^{2H}$
\citep{Stu98}, so that the mean variation over distance
$\lambda$ is $\Delta_I(\lambda)\varpropto\lambda^{H}$.

To generate a three-dimensional turbulent fractal cloud with a
log-normal density distribution we follow \citet{Elm02}; that is,
the Gaussian intensity distribution $I_{i,j,k}$ is exponentiated to
generate a density distribution with a log-normal PDF. The masses
$m_{i,j,k}$ in the lattice are $m_{i,j,k}=\emmax \, exp[\alpha \, \,
 (\frac{I_{i,j,k}}{I_{max}}-1)]$, where $I_{max}$ is the maximum
value of intensities $I_{i,j,k}$ and $\alpha$ is the contrast factor.
Since for a given $\alpha$ the intensity $I_{i,j,k}$ is proportional
to $ln(m_{i,j,k})$, the mean variation of $ln(m)$ over distance
$\lambda$ is proportional to $\lambda^{H}$. That is, small values of
$H$ produce very rough structures, whereas very smooth structures are
produced with $H$ close to 1. \citet{Elm02} restrict their results
to the case $H=1/3$.

By construction the densest voxel has a mass $\emmax$. The total mass
in the lattice $M_{cl}$, and therefore the mean mass per voxel ${\bar
 m}$, depends on $\alpha$ and $H$. However, due to the finite size of
the lattice, two realizations using the same parameter values $\alpha$
and $H$ but different randomization of the phases in general do not
contain the same total mass $M_{cl}$. Also, due to the finite size of
the array, the density PDF departs from the log-normal form,
especially at low densities and large values of $H$. To compare the
results among fBm clouds we adjust the value of $\alpha$ in each
realization in order to have the same mass $M_{cl}$ and the same
maximum voxel mass $\emmax$ in the lattice. 

\begin{figure}
\epsscale{0.9} \plotone{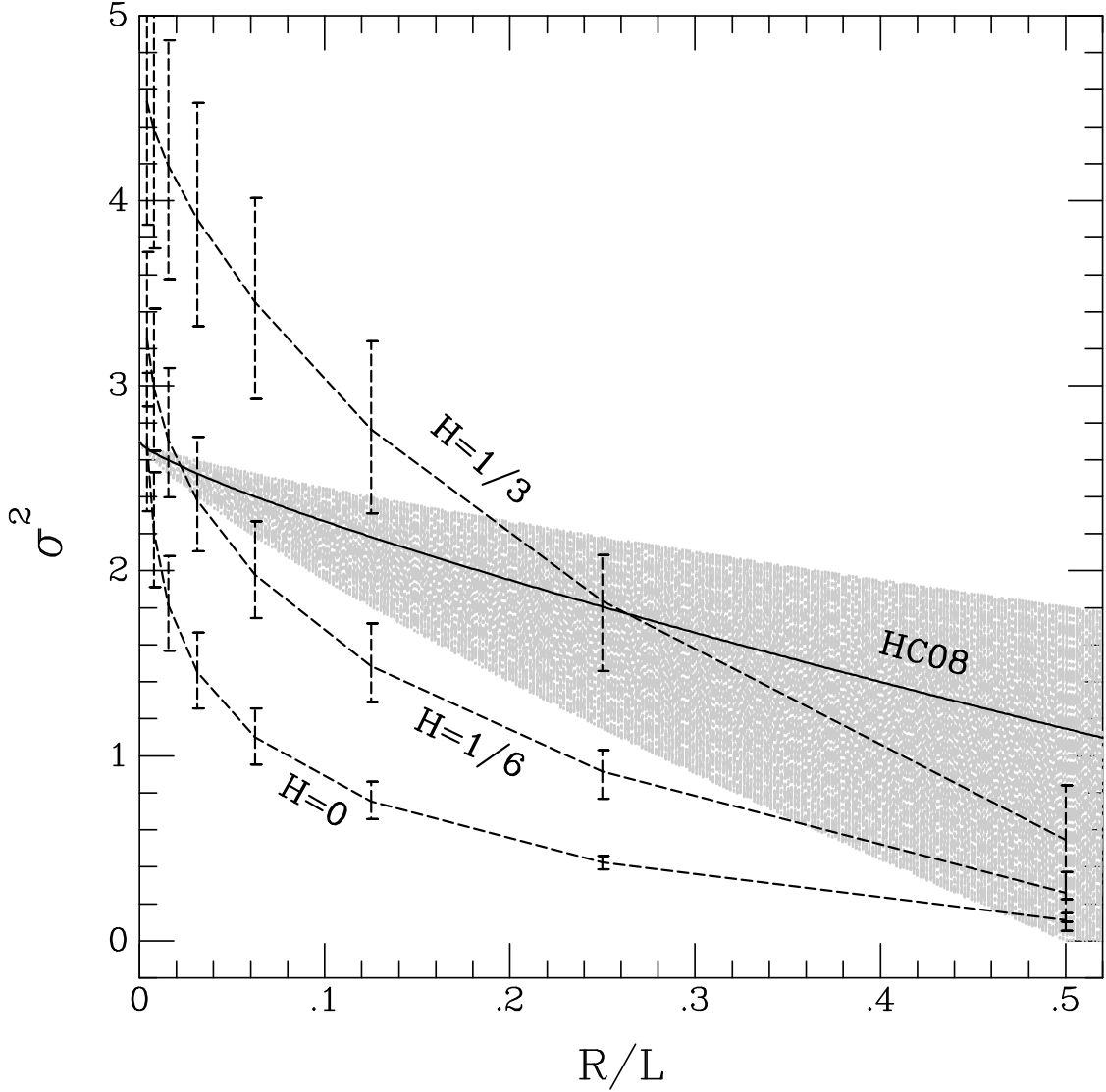}
\caption
{Dependence of the standard deviation on the smoothing scale
 $\sigma^2(R)$ in fBm clouds with three different values of the Hurst
 parameter $H$. Each curve with error bars corresponds to the average
 value for ten fBm clouds with the same $H$. The dashed area
 represents the HC08 values of $\sigma^2$ given by eq. (\ref{sig2_R})
 with an injection scale length in the range $L/2 \leq L_i \leq 2 L$.
 The curve with the label HC08 corresponds to eq. (\ref{sig2_R}) with
 $L_i=L$ and $\sigma_0$ calculated from eqs. (\ref{mach-cloud} -
 \ref{s2-cloud}). The parameters used are $N_{vox}=2^8$,
 $\emmax=0.07\, \emsun$, $T=10$K, $\mu=2.33$ and $\eta=0.4$, which
 correspond to $L=2.7$ pc, $M_{cl}=860 \, \emsun$, $\sigma_0=1.64$,
 as in the case of the third line of Table 1. }
\label{s_VS_R}
\end{figure}

Figure (\ref{s_VS_R}) shows $\sigma^2(R/L)$ for fBm clouds with
various values of the Hurst exponent $H$. Each curve represents the
average over 10 realizations and the error bars indicate the 10 and 90
percentiles. As in Fig. \ref{s_VS_R_c_r} the cloud mass $M_{cl}$ and
the value of $\sigma_0$ in the HC08 theory (eq.\ref{sig2_R}) are the
values corresponding to the case in the third line of Table 1. The
dashed area represents the values of $\sigma^2$ given by
eq.(\ref{sig2_R}) with an injection scale length in the range $L/2\leq
L_i \leq 2 L$.

\begin{figure}
\epsscale{0.9} \plotone{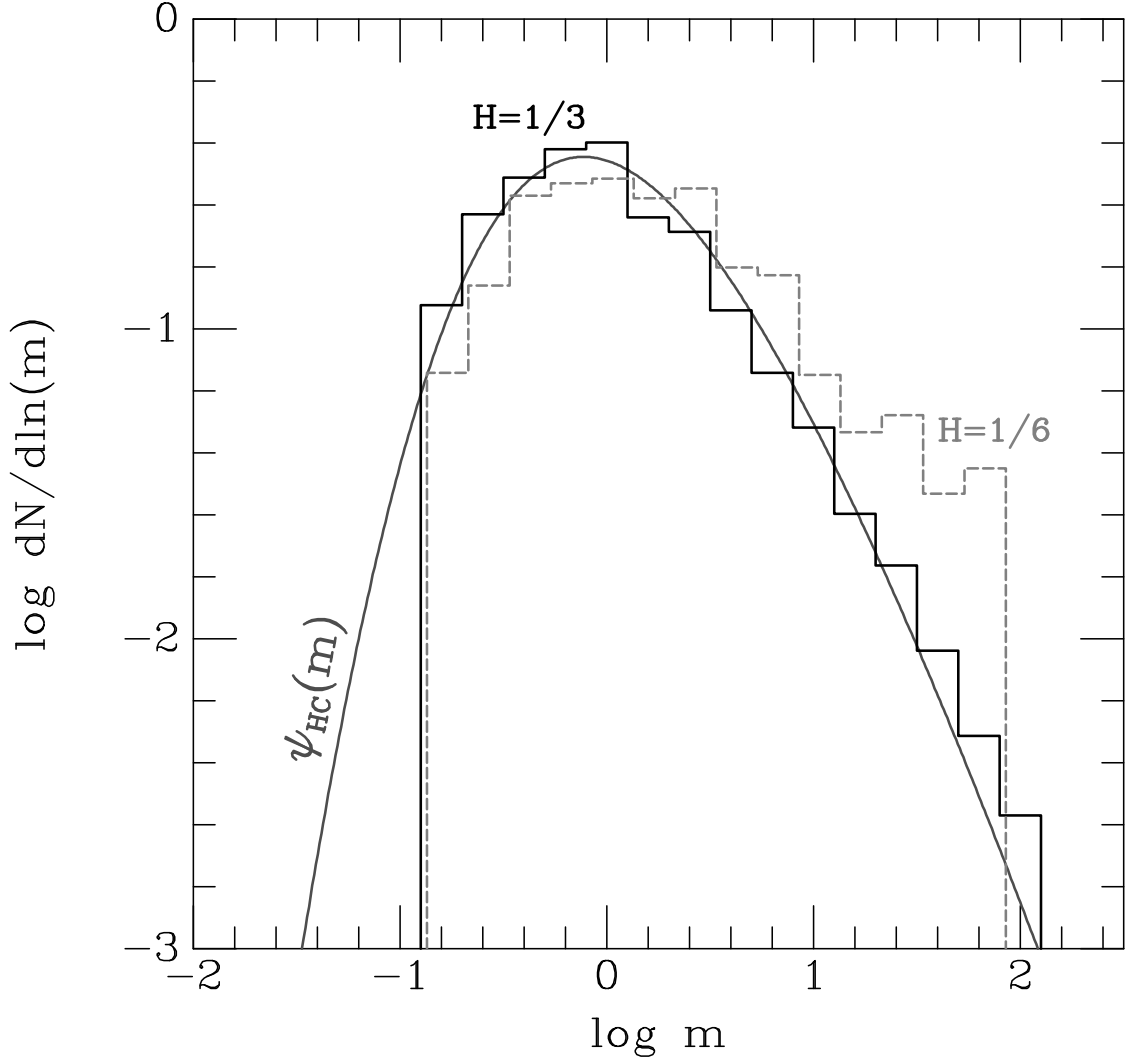}
\caption
{Mass distribution of the cores formed in a total of 20 different
 simulations of fBm clouds. The solid (dashed) line histogram
 corresponds to the mass function of fBm clouds with $H=1/3$
 ($H=1/6$). The parameter values in all simulations are
 $N_{vox}=2^8$, $\emmax=0.07 \, \emsun$, $T=10$ K and $\mu=2.33$. The
 contrast factor $\alpha$ in each simulation is adjusted to produce a
 cloud with a mass of 860 \msun that corresponds to these parameter
 values (see third line in Table1). As in
 Fig. (\ref{fig:corner-hc08}), the sharp rise at core mass of about
 0.1 \msun is due to the adopted $\emmax$ value. The continuous
 curve shows $\psi_{\rm HC}(m)$ for these parameter values. }
\label{fig:mf-h1o6y1o3-20realiz}
\end{figure}

Figure (\ref{fig:mf-h1o6y1o3-20realiz}) shows the mass distribution of
the cores formed in 20 different simulations with the same parameters
$H$ and $\emmax$, and the contrast factor $\alpha$ adjusted to always
produce the same cloud mass. The agreement with $\psi_{\rm HC}(m)$ is
much better for the case $H=1/3$ than for $H=1/6$. When $H=1/6$ there
is a clear excess of high mass cores compared to $\psi_{\rm HC}(m)$,
which can be understood in terms of the dependence $\sigma^2$ on
$R$. As shown in Fig. \ref{s_VS_R}, except for $R/L\lesssim 0.03$ the
curve $\sigma^2(R;H=1/6)$ is below the HC08 curve
$\sigma^2_0(1-(R/L)^{2\eta})$. When $\sigma^2(R)$ is small, the
density PDF is a narrow distribution around ${\bar \rho}$ and at these
densities and sizes only massive cores can form. Note that the curve
$\sigma^2(R;H=1/3)$ intersects the HC08 curve at a larger scale and
that is why a fBm cloud with $H=1/3$ produces in general more low-mass
cores than that predicted by the HC08 theory. These results highlight
the importance of the spatial distribution of the gas on the resulting
mass distribution of collapsing objects.

\begin{table}
\begin{center}
\begin{tabular}{c}
\hspace{0.0cm}TABLE 2\\
\hspace{0.0cm} Average properties of the cores in fBm clouds\\
\end{tabular}\\
\begin{tabular}{ccccc}
\hline \hline $H$& ${\cal N}_{cores}$\tablenotemark{a}& ${\bar
 m}_{core}$\tablenotemark{b}& $F_{m,c/g}$\tablenotemark{c}&
$F_{h,{\rm ms}}$\tablenotemark{d}\\ \hline 0  & $  28\pm  6$ &
$ 16.28\pm 4.90$ & $ 0.49\pm 0.06$ & $0.13\pm0.05$ \\ 1/6 & $
89\pm  25$ & $ 5.07\pm 1.96$ & $ 0.47\pm 0.06$ &
$0.022\pm0.02$ \\ 1/5 & $ 101\pm  29$ & $ 4.47\pm 1.78$ & $
0.47\pm 0.07$ & $0.017\pm0.013$ \\ 1/4 & $ 132\pm  38$ & $
3.43\pm 1.43$ & $ 0.47\pm 0.06$ & $0.009\pm0.009$ \\ 1/3 & $
201\pm  57$ & $ 2.42\pm 0.87$ & $ 0.51\pm 0.06$ &
$0.005\pm0.005$ \\ 1/2 & $ 405\pm 110$ & $ 1.31\pm 0.42$ & $
0.57\pm 0.06$ & $0.0006\pm0.0015$ \\ 3/4 & $ 797\pm 177$ & $
0.76\pm 0.17$ & $ 0.67\pm 0.05$ & $0.0000\pm0.0000$ \\ \hline
\end{tabular}
\tablenotetext{}{For each value of $H$, the values quoted correspond
 to the average over 20 simulations of fBm clouds with different
 random phase but identical parameter values $N_{vox}=2^8$,
 $\emmax=0.07\, \emsun$, $T=10$K, $\mu=2.33$ and $\eta=0.4$, that
 correspond to $L=2.7$ pc, $M_{cl}=860 \, \emsun$, $\sigma_0=1.64$,
 as in the third case in Table 1. The errors indicate the standard
 deviation of the 20 values of each parameter for each value of $H$}
\tablenotetext{a} {Number of cores per cloud.} \tablenotetext{b}
       {Average mass of cores in a cloud in solar masses.}
       \tablenotetext{c} {Fraction of the mass of the cloud
        that collapses in cores.} \tablenotetext{d} {Fraction
        of stars with masses over 8 $\emsun$ estimated with
        eq. (\ref{Fh}).}
\end{center}
\end{table}

A complex interplay of physical processes determines the final mass of
the stars that form from a particular configuration of cores in a
star-forming region. However, the scaling observed between the CMF and
the stellar IMF indicate that some average relations between these two
distributions can be established. For example, the fraction
$F_{h,{\rm ms}}$ of individual main-sequence stars that will end as
core collapse supernovae or the mean mass ${\bar m_*}$ of stars in the
IMF can be estimated in terms of core to stellar system efficiency
($\epsilon\sim 1/3$, Matzner \& McKee 2000), and the dependence of the
binary fraction on the system mass \citep{Lad06}. 

The stellar mean mass can be approximated as ${\bar m_*}\approx
(\epsilon/R_s) {\bar m}_{core}$, where $R_s$ is the mean number of
stars per system in the IMF. In other words, $R_s$ is defined as the
ratio of the total number of stars to the total number of systems
(single star systems + multiple star systems). \citet{Par11}
estimate that the mean mass of the objects in the individual star IMF
is ${\bar m_*}\simeq 0.75 \emsun$, so that, assuming that $R_s\simeq
1.3$ \citep{Lad06}, the corresponding mean core mass is ${\bar m}_{core}
\approx 3 \emsun$. Note that this estimate of ${\bar m}_{core}$
assumes that the core to star efficiency does not depend on the mass
of the core or on the number of stellar objects formed.

The fraction of individual main-sequence stars formed with masses over
$m_h\sim 8 \, \emsun$ can be estimated as \beq F_{h,{\rm ms}}\simeq
\frac {R_{s,h} \, {\cal N}(m_{core} > R_{s,h} \, m_h /\epsilon)}
  {R_{s} \, {\cal N}(m_{core} > m_{bd}/\epsilon)},
\label{Fh}
\eeq where $R_{s,h}$ is the mean number of high-mass stars ($m>m_h$)
formed in high mass cores $(m_{core} > R_{s,h} \, m_h
/\epsilon)$. The values quoted in Table 2 correspond to $R_s\simeq
1.3$ and $R_{s,h}\simeq 2$. Note that eq.(\ref{Fh}) assumes that the
stellar system formed in a core does not disaggregate. Additionally,
eq.(\ref{Fh}) neglects the high-mass primary stars in systems with
companions having masses below $m_h$. However, the error introduced
is small because most high mass stars have a companion of similar mass
(Ma{\'{\i}}z Apell{\'a}niz 2008). Eq.(\ref{Fh}) also neglects very
low-mass primaries with brown dwarf companions ($m<m_{bd}\simeq 0.08\,
\emsun$). Nonetheless, since the binary fraction of very low-mass
stars is small $\sim 0.2$ \citep{Rei06,Bur07},
the vast majority of systems with $m > m_{bd}$ have main sequence
primaries. \citet{Par11} estimate that $F_{h,{\rm
 ms}}\simeq (7 - 8) \times 10^{-3}$.

\begin{figure}
\epsscale{0.9} \plotone{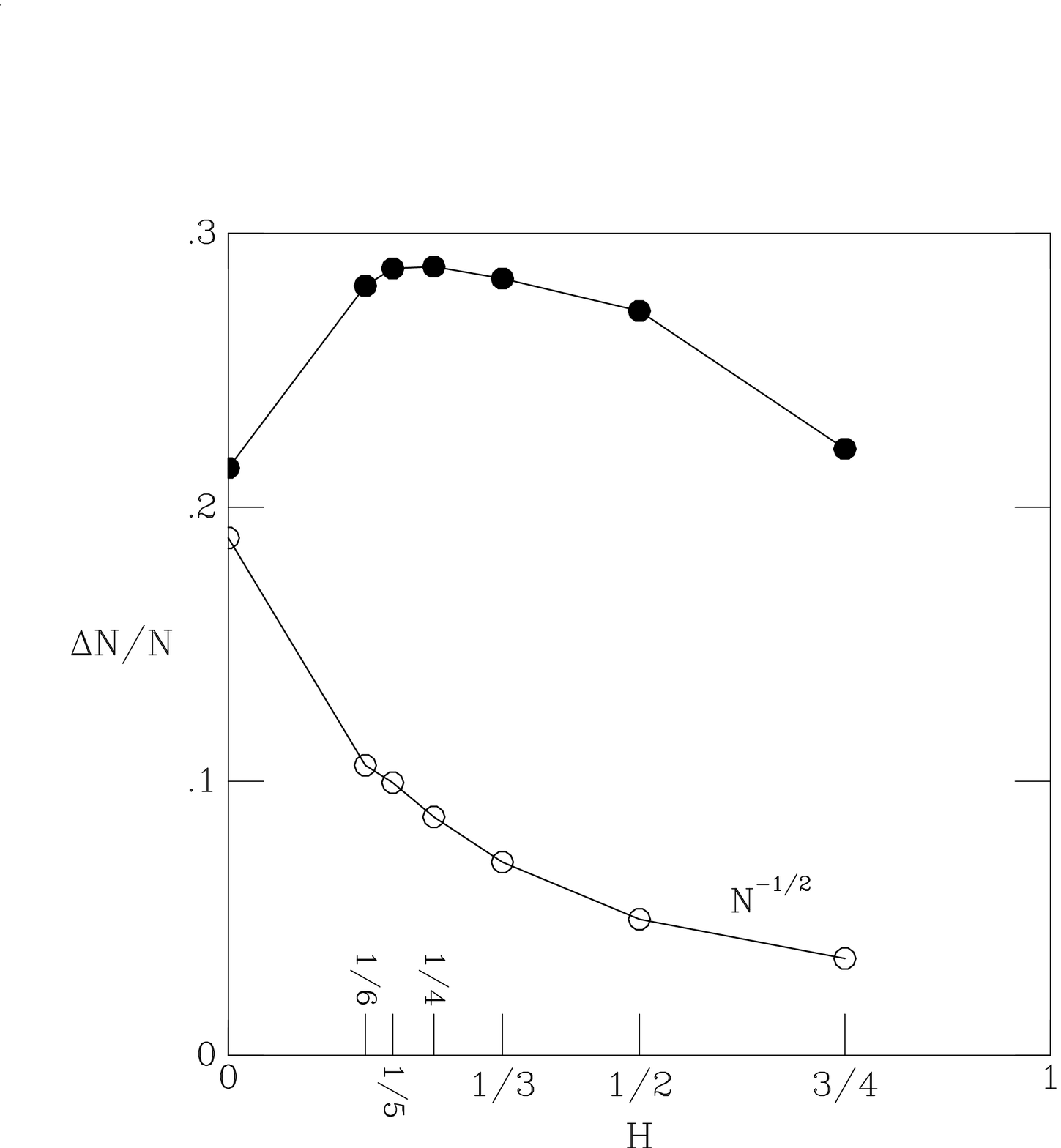}
\caption
{Relative variation of the number of cores ($\Delta{\cal
 N}_{cores}/{\cal N}_{cores}$) in 20 simulations as function of the
 Hurst parameter $H$. The open circles are the corresponding values
 $1/\sqrt{\cal N}$ expected if they were statistical variations of
 the number of cores. }
\label{fig:varcore_H}
\end{figure}

For seven values of the Hurst exponent $H$, Table 2 summarizes the
average properties of the cores formed in sets of 20 simulations of
fBm clouds with the same mass $M_{cl}$ but a different setup of the
random Fourier phases. For the parameter values used for the
simulations in Table 2, fBm clouds with $1/4 \lesssim H \lesssim 1/3$
produce an average core mass distribution that agrees both with the
theoretical HC08 CMF and with the expected values of ${\bar
 m}_{core}\approx 3 \emsun$ and $F_{h,{\rm ms}}\simeq (7 - 8) \times
10^{-3}$. Note also that as $H$ increases the number of cores
increases and their average mass decreases. Figure
(\ref{fig:varcore_H}) shows the dependence of the relative variation
of the number of cores $\Delta{\cal N}_{cores}/{\cal N}_{cores}$ as
function of $H$, where $\Delta{\cal N}_{cores}$ is the standard
deviation of ${\cal N}_{cores}$ in the 20 simulations. Except for
$H=0$ the variations in the number of cores largely exceed the
expected $\sqrt{\cal N}$ statistical variations. The analysis of the
dependence of these variations on the considered physical processes
and cloud structure is out of the scope of the present
study. However,we notice here that for a fixed volume of simulation
and a constant cloud mass the relative variation DeltaN/N is about
constant for fBm clouds with H values in the range considered in
Fig. 5.

\subsection{Spatial distribution of cores}

\begin{figure}
\epsscale{0.9} \plotone{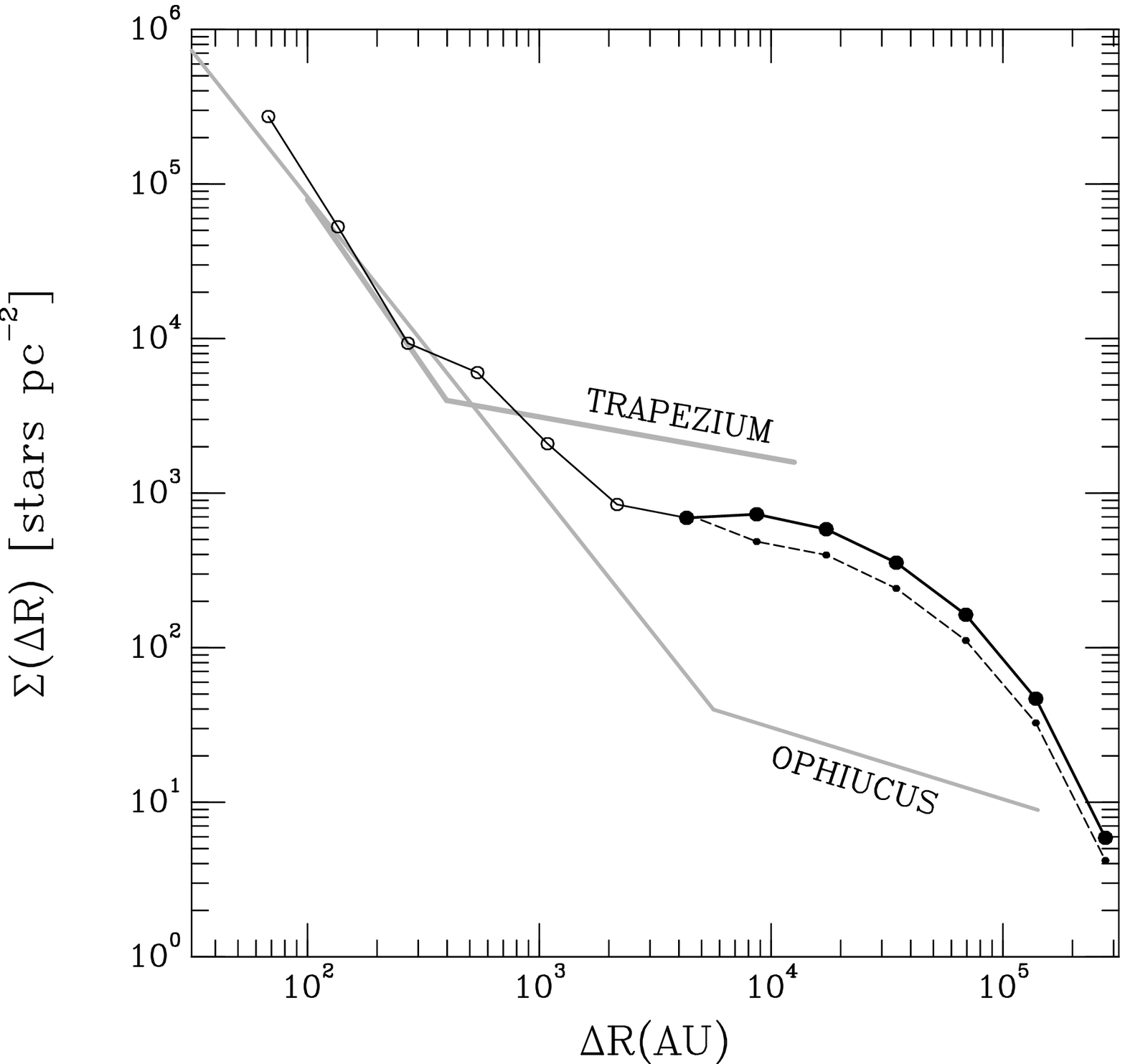}
\caption
{Surface density of companions $\Sigma(\Delta R)$ vs the separation in
 AU for a single simulation of fBm clouds with H=1/3. The small dots
 connected by dashed lines correspond to $\Sigma(\Delta R)$ for the
 cores. The large dots connected by solid lines correspond to the
 surface density of companions systems ($\Delta R > l_{vox}$) when
 half of the cores are assumed to fragment into two units. The open
 circles connected by light lines correspond to the surface density
 of companion stars assuming that the binary separation $\Delta R$ is
 less than $l_{vox}$ with a probability distribution $p(R)\propto
 \Delta R^{-1/2}$. The two segment power-law gray lines represent
 the surface density of companions for the Orion Trapezium and
 Ophiuchus star formation regions \citep{Sim97}. }
\label{fig:sdc_2D}
\end{figure}

The spatial distribution of the collapsing cores can be characterized
by the surface density of companions (SDC) measured on a 2d projection
of the positions of the cores by sampling all pair of cores over bins
of separation $\Delta R$. In order to compare with the SDC in young
clusters (Simon 1997) we assume that half of the cores fragment to
form two stars with 3D separations $\Delta R\leq l_{vox}$ following a
probability distribution $p(R)\propto \Delta R^{-1/2}$. Figure
(\ref{fig:sdc_2D}) shows the surface density of companions for a
single simulation of a fBm cloud with H=1/3 and the same parameter
values used in Fig. (\ref{fig:mf-h1o6y1o3-20realiz}) and the case
H=1/3 in Table 2. The labeled power-laws in Fig. (\ref{fig:sdc_2D})
are the SDCs reported by \citet{Sim97} for the Orion Trapezium star
formation region and for the Ophiuchus star formation region. The SDC
for this particular simulation is in between the observed SDC's for
these two regions. The fall of the SDC at large radii is due to edge
effects when $\Delta R$ is of the order of size $L$ of the simulation.
For other simulations with the same parameter values the SDCs are
similar, but when the parameter H is increased the SDC curves shift
upwards. 

\section{Conclusions}
We have proposed a procedure to obtain the prestellar core mass
distribution that results from the collapse of prestellar cores in
clumps by assuming that the densest regions collapse first and form
the smaller objects. At small scales, thermal support dominates and
determines the mass distribution of cores at low masses, whereas at
the largest scales turbulence dominates the support and determines
themass distribution at high masses. The numerical method proposed
here make use of a small number of parameters, namely $N_{vox}, T,
\mu, \eta$, together with the cloud properties (i.e. $M_{cl}$ and
$H$) and the assumptions that the mass in the densest voxel $m_{max}$
is equal to the Jeans mass and that the cloud as a whole is marginally
stable. When the proposed method is applied to a mass distribution
whose density PDF is a log-normal at all smoothing scales $R$ and its
standard deviation $\sigma(R)$ is given by eq. (\ref{sig2_R}), the
average mass distribution agrees with the CMF predicted by the
analytical theory of the IMF proposed by HC08. The HCS method can be
seen as a numerical version of the HC08 theory, and there is univocal
correspondence between the parameters in both models. 

Both the Padoan-Nordlund IMF and the Hennebelle-Chabrier IMF apply to
particular star-forming cloud conditions. In order to determine an
average IMF that can be compared with observations of stars from
different clouds, it is necessary to average their theoretical IMFs
for a distribution of cloud temperatures, densities and Mach numbers.
The core mass distribution from our method is even more dependent on
cloud property since, as we have shown, the masses of the resulting
cores also depend on the particular distribution of mass within the
cloud. Large variations in the resulting core mass distribution are
observed in fBm clouds with the same mass $M_{cl}$ and Hurst exponent
$H$, but a different setup of the random Fourier phases. As shown in
Table 2 and Fig. (\ref{fig:varcore_H}), the number of cores in a set
of 20 simulations display variations that largely exceed the expected
$\sqrt{\cal N}$ statistical variations. Due to its simplicity the HCS
method is computationally efficient at obtaining the mass and position
of the cores that collapse in an arbitrary distribution of
gas. Therefore the HCS method is well suited to analyzing the effects
produced by changes in the physics over a large number of initial
conditions.

We have applied the HCS method to lattices with a number of cells up
to $2^{8\times 3}$, which represent clumps of mass $\sim 10^3 \,
\emsun$ and size $\sim 3$ pc, but larger lattices can be
processed. There is no restriction in the way the mass in the voxels
is assigned, but we have focused on fBm clouds that have been used as
analogs of real interstellar clouds. We confirm that fBm clouds with
$H\simeq1/3$, corresponding to $\gamma=11/3$ (Elmegreen 2002), give
better agreement with the theoretical CMF derived by Hennebelle and
Chabrier and the observed IMF. We have also shown that the spatial
distribution of the cores for fBm clouds with $H=1/3$ has a surface
density of companions that resembles that of young stellar clusters
(Simon 1997). Since the HCS method provides the sequence and location
of newly formed stars, the method can be easily modified to consider
radiative feedback effects. 

\acknowledgments

We wish to thank David Hollenbach and Christopher McKee for their
valuable comments. AP was supported as guest researcher for part of
this research by the IESA-CSIC. N.S. acknowledges financial support
from the Ministerio de Econom\'ia y Competitividad of Spain through
grant AYA2011-29754-C03-01, and EJA through grant AYA2010-17631.

\end{document}